\input harvmac
\overfullrule=0pt
\parindent 25pt
\tolerance=10000
\input epsf

\newcount\figno
\figno=0
\def\fig#1#2#3{
\par\begingroup\parindent=0pt\leftskip=1cm\rightskip=1cm\parindent=0pt
\baselineskip=11pt
\global\advance\figno by 1
\midinsert
\epsfxsize=#3
\centerline{\epsfbox{#2}}
\vskip 12pt
{\bf Fig.\ \the\figno: } #1\par
\endinsert\endgroup\par
}
\def\figlabel#1{\xdef#1{\the\figno}}
\def\encadremath#1{\vbox{\hrule\hbox{\vrule\kern8pt\vbox{\kern8pt
\hbox{$\displaystyle #1$}\kern8pt}
\kern8pt\vrule}\hrule}}

 \def\ep{{\epsilon}}

 \def\T{{\Theta}}
 \def\frac#1#2{{#1\over #2}}

 \def\s{\sqrt}

 \def\al{\alpha'}
 \def\de{\partial}

 \def\f {\frac}
 \def\ti{\tilde}
 \def\ap{\alpha}

 \def\la{\langle}
 \def\lb{\rangle}
 \def\ep{\epsilon}

 \def\vp{\varphi}

\lref\mopl{ G.~W.~Moore and R.~Plesser, ``Classical scattering in
(1+1)-dimensional string theory,'' Phys.\ Rev.\ D {\bf 46}, 1730
(1992) [arXiv:hep-th/9203060].
}

\lref\PoS{ J.~Polchinski, ``Classical limit of (1+1)-dimensional
string theory,'' Nucl.\ Phys.\ B {\bf 362}, 125 (1991).
}

\lref\Al{ S.~Alexandrov, ``Backgrounds of 2D string theory from
matrix model,'' arXiv:hep-th/0303190.
}

\lref\KSone{ J.~L.~Karczmarek and A.~Strominger, ``Matrix
cosmology,'' JHEP {\bf 0404}, 055 (2004) [arXiv:hep-th/0309138].
}

\lref\KStwo{ J.~L.~Karczmarek and A.~Strominger, ``Closed string
tachyon condensation at c = 1,'' JHEP {\bf 0405}, 062 (2004)
[arXiv:hep-th/0403169];
J.~L.~Karczmarek, A.~Maloney and A.~Strominger,
``Hartle-Hawking vacuum for c = 1 tachyon condensation,''
arXiv:hep-th/0405092;
M.~Ernebjerg, J.~L.~Karczmarek and J.~M.~Lapan,
``Collective field description of matrix cosmologies,''
JHEP {\bf 0409}, 065 (2004)
[arXiv:hep-th/0405187].
}

\lref\DDLM{ S.~R.~Das, J.~L.~Davis, F.~Larsen and P.~Mukhopadhyay,
``Particle production in matrix cosmology,'' Phys.\ Rev.\ D {\bf
70}, 044017 (2004) [arXiv:hep-th/0403275];
P.~Mukhopadhyay, ``On the problem of particle production in c = 1
matrix model,'' JHEP {\bf 0408}, 032 (2004)
[arXiv:hep-th/0406029].
}

\lref\MPY{ D.~Minic, J.~Polchinski and Z.~Yang, ``Translation
invariant backgrounds in (1+1)-dimensional string theory,'' Nucl.\
Phys.\ B {\bf 369}, 324 (1992).
}

\lref\AKKT{ S.~Y.~Alexandrov, V.~A.~Kazakov and I.~K.~Kostov,
``Time-dependent backgrounds of 2D string theory,'' Nucl.\ Phys.\
B {\bf 640}, 119 (2002) [arXiv:hep-th/0205079].
}

\lref\DM{
B.~C.~Da Cunha and E.~J.~Martinec,
``Closed string tachyon condensation and worldsheet inflation,''
Phys.\ Rev.\ D {\bf 68}, 063502 (2003)
[arXiv:hep-th/0303087].
}

\lref\tdsa{ D.~J.~Gross and N.~Miljkovic, ``A Nonperturbative
Solution Of D = 1 String Theory,'' Phys.\ Lett.\ B {\bf 238}, 217
(1990);
}

\lref\tdsb{ E.~Brezin, V.~A.~Kazakov and A.~B.~Zamolodchikov,
``Scaling Violation In A Field Theory Of Closed Strings In One
Physical Dimension,'' Nucl.\ Phys.\ B {\bf 338}, 673 (1990);
}

\lref\tdsc{ P.~Ginsparg and J.~Zinn-Justin, ``2-D Gravity + 1-D
Matter,'' Phys.\ Lett.\ B {\bf 240}, 333 (1990).
}

\lref\DK{P.~Di Francesco and D.~Kutasov, ``Correlation functions
in 2-D string theory,'' Phys.\ Lett.\ B {\bf 261}, 385 (1991);
``World sheet and space-time physics in two-dimensional
(Super)string theory,'' Nucl.\ Phys.\ B {\bf 375}, 119 (1992)
[arXiv:hep-th/9109005].
}

\lref\GTT{ S.~Gukov, T.~Takayanagi and N.~Toumbas, ``Flux
backgrounds in 2D string theory,'' JHEP {\bf 0403}, 017 (2004)
[arXiv:hep-th/0312208].
}

\lref\MV{ J.~McGreevy and H.~Verlinde, ``Strings from tachyons:
The c = 1 matrix reloaded,'' JHEP {\bf 0312}, 054 (2003)
[arXiv:hep-th/0304224].
}

\lref\KMS{ I.~R.~Klebanov, J.~Maldacena and N.~Seiberg, ``D-brane
decay in two-dimensional string theory,'' JHEP {\bf 0307}, 045
(2003) [arXiv:hep-th/0305159].
}

\lref\TT{T.~Takayanagi and N.~Toumbas, ``A matrix model dual of
type 0B string theory in two dimensions,'' JHEP {\bf 0307}, 064
(2003) [arXiv:hep-th/0307083].
}

\lref\KlR{I.~R.~Klebanov, ``String theory in two-dimensions,''
[arXiv:hep-th/9108019].}

\lref\six{M.~R.~Douglas, I.~R.~Klebanov, D.~Kutasov, J.~Maldacena,
E.~Martinec and N.~Seiberg, ``A new hat for the c = 1 matrix
model,'' [arXiv:hep-th/0307195].
}

\lref\ADKMV{ M.~Aganagic, R.~Dijkgraaf, A.~Klemm, M.~Marino and
C.~Vafa, ``Topological strings and integrable hierarchies,''
arXiv:hep-th/0312085.
}

\lref\Se{ N.~Seiberg, ``Notes On Quantum Liouville Theory And
Quantum Gravity,'' Prog.\ Theor.\ Phys.\ Suppl.\  {\bf 102}, 319
(1990).
}

\lref\INOS{
H.~Ita, H.~Nieder, Y.~Oz and T.~Sakai,
``Topological B-model, matrix models, c-hat = 1 strings
and quiver gauge
theories,''
JHEP {\bf 0405}, 058 (2004)
[arXiv:hep-th/0403256].
}

\lref\DOV{
U.~H.~Danielsson, M.~E.~Olsson and M.~Vonk,
``Matrix models, 4D black holes and topological strings
on non-compact
Calabi-Yau manifolds,''
arXiv:hep-th/0410141.
}

\lref\KMSG{
D.~Kutasov, E.~J.~Martinec and N.~Seiberg,
``Ground rings and their modules in 2-D gravity
with $c \leq 1$ matter,''
Phys.\ Lett.\ B {\bf 276}, 437 (1992)
[arXiv:hep-th/9111048].
}

\lref\GOV{ S.~Govindarajan, T.~Jayaraman and V.~John, ``Chiral
rings and physical states in c $<$ 1 string theory,'' Nucl.\
Phys.\ B {\bf 402}, 118 (1993) [arXiv:hep-th/9207109].
}

\lref\AJ{ J.~Avan and A.~Jevicki, ``Classical integrability and
higher symmetries of collective string field theory,'' Phys.\
Lett.\ B {\bf 266}, 35 (1991).
}

\lref\TaD{
T.~Takayanagi,
``Notes on D-branes in 2D type 0 string theory,''
JHEP {\bf 0405}, 063 (2004)
[arXiv:hep-th/0402196].
}

\lref\GM{
P.~H.~Ginsparg and G.~W.~Moore,
``Lectures on 2-D gravity and 2-D string theory,''
arXiv:hep-th/9304011.
}

\lref\PoR{
J.~Polchinski,
``What is string theory?,''
arXiv:hep-th/9411028.
}

\lref\Nak{
Y.~Nakayama,
``Liouville field theory: A decade after the revolution,''
arXiv:hep-th/0402009.
}

\lref\TS{ T.~Takayanagi, ``Comments on 2D type IIA string and
matrix model,'' arXiv:hep-th/0408086.
}

\lref\COL{A.~B.~Zamolodchikov and A.~B.~Zamolodchikov, ``Structure
constants and conformal bootstrap in Liouville field theory,''
Nucl.\ Phys.\ B {\bf 477}, 577 (1996) [arXiv:hep-th/9506136].
}

\lref\DO{ H.~Dorn and H.~J.~Otto, ``Two and three point functions
in Liouville theory,'' Nucl.\ Phys.\ B {\bf 429}, 375 (1994)
[arXiv:hep-th/9403141].
}

\lref\COR{ J.~Teschner, ``Remarks on Liouville theory with
boundary,'' arXiv:hep-th/0009138.
}

\lref\Ko{ I.~K.~Kostov, ``String equation for string theory on a
circle,'' Nucl.\ Phys.\ B {\bf 624}, 146 (2002)
[arXiv:hep-th/0107247];
``Integrable flows in c = 1 string theory,'' J.\ Phys.\ A {\bf
36}, 3153 (2003) [Annales Henri Poincare {\bf 4}, S825 (2003)]
[arXiv:hep-th/0208034].
}

\lref\EK{T.~Eguchi and H.~Kanno, ``Toda lattice hierarchy and the
topological description of the c = 1 string theory,'' Phys.\
Lett.\ B {\bf 331}, 330 (1994) [arXiv:hep-th/9404056].
}

\lref\KKK{ V.~Kazakov, I.~K.~Kostov and D.~Kutasov, ``A matrix
model for the two-dimensional black hole,'' Nucl.\ Phys.\ B {\bf
622}, 141 (2002) [arXiv:hep-th/0101011].
}

\lref\ST{ A.~Strominger and T.~Takayanagi, ``Correlators in
timelike bulk Liouville theory,'' Adv.\ Theor.\ Math.\ Phys.\
{\bf 7}, 369 (2003) [arXiv:hep-th/0303221].
}

\lref\SC{V.~Schomerus, ``Rolling tachyons from Liouville theory,''
JHEP {\bf 0311}, 043 (2003) [arXiv:hep-th/0306026].
}

\lref\ZZ{ A.~B.~Zamolodchikov and A.~B.~Zamolodchikov, ``Liouville
field theory on a pseudosphere,'' arXiv:hep-th/0101152.
}

\lref\FZZ{V.~Fateev, A.~B.~Zamolodchikov and A.~B.~Zamolodchikov,
``Boundary Liouville field theory. I: Boundary state and boundary
two-point function,'' arXiv:hep-th/0001012.
}

\lref\T{ J.~Teschner, ``Remarks on Liouville theory with
boundary,'' arXiv:hep-th/0009138.
}

\lref\STL{A.~Strominger, ``Open string creation by S-branes,''
arXiv:hep-th/0209090.
}

\lref\LNT{ F.~Larsen, A.~Naqvi and S.~Terashima, ``Rolling
tachyons and decaying branes,'' JHEP {\bf 0302}, 039 (2003)
[arXiv:hep-th/0212248].
}

\lref\GS{M.~Gutperle and A.~Strominger, ``Timelike boundary
Liouville theory,'' Phys.\ Rev.\ D {\bf 67}, 126002 (2003)
[arXiv:hep-th/0301038].
}

\lref\LLM{ N.~Lambert, H.~Liu and J.~Maldacena, ``Closed strings
from decaying D-branes,'' arXiv:hep-th/0303139.
}

\lref\Sen{ A.~Sen, ``Rolling tachyon,'' JHEP {\bf 0204}, 048
(2002) [arXiv:hep-th/0203211].
}

\lref\SCT{S.~Fredenhagen and V.~Schomerus, ``Boundary Liouville
theory at c = 1,'' arXiv:hep-th/0409256.
}

\lref\DJ{S.~R.~Das and A.~Jevicki, ``String Field Theory And
Physical Interpretation Of D = 1 Strings,'' Mod.\ Phys.\ Lett.\ A
{\bf 5}, 1639 (1990).
}

\lref\HT{ Y.~Hikida and T.~Takayanagi, ``On solvable
time-dependent model and rolling closed string tachyon,''
arXiv:hep-th/0408124.
}

\lref\Ka{ A.~Kapustin, ``Noncritical superstrings in a
Ramond-Ramond background,'' JHEP {\bf 0406}, 024 (2004)
[arXiv:hep-th/0308119].
}

\lref\JY{A.~Jevicki and T.~Yoneya, ``A Deformed matrix model and
the black hole background in two-dimensional
Nucl.\ Phys.\ B {\bf 411}, 64 (1994) [arXiv:hep-th/9305109].
}

\lref\Mac{ G.~W.~Moore, N.~Seiberg and M.~Staudacher, ``From loops
to states in 2-D quantum gravity,'' Nucl.\ Phys.\ B {\bf 362}, 665
(1991);
G.~W.~Moore and N.~Seiberg, ``From loops to fields in 2-D quantum
gravity,'' Int.\ J.\ Mod.\ Phys.\ A {\bf 7}, 2601 (1992).
}

\lref\Ma{ E.~J.~Martinec, ``The annular report on non-critical
string theory,'' arXiv:hep-th/0305148.
}

\lref\MaR{ E.~J.~Martinec, ``Matrix models and 2D string theory,''
arXiv:hep-th/0410136.
}

\lref\witteng{
E.~Witten,
``Ground ring of two-dimensional string theory,''
Nucl.\ Phys.\ B {\bf 373}, 187 (1992)
[arXiv:hep-th/9108004].
}

\lref\GV{
D.~Ghoshal and C.~Vafa,
``C = 1 string as the topological theory of the conifold,''
Nucl.\ Phys.\ B {\bf 453}, 121 (1995)
[arXiv:hep-th/9506122].
}

\lref\DMP{
R.~Dijkgraaf, G.~W.~Moore and R.~Plesser,
``The Partition function of 2-D string theory,''
Nucl.\ Phys.\ B {\bf 394}, 356 (1993)
[arXiv:hep-th/9208031].
}

\lref\SH{
N.~Seiberg and D.~Shih,
``Branes, rings and matrix models in minimal (super)string theory,''
JHEP {\bf 0402}, 021 (2004)
[arXiv:hep-th/0312170].
}

\lref\IM{
N.~Itzhaki and J.~McGreevy,
``The large N harmonic oscillator as a string theory,''
arXiv:hep-th/0408180.
}

\lref\SenC{ A.~Sen, ``Rolling tachyon boundary state, conserved
charges and two dimensional string theory,'' JHEP {\bf 0405}, 076
(2004) [arXiv:hep-th/0402157];
``Symmetries, conserved charges and (black) holes in two
dimensional string theory,'' arXiv:hep-th/0408064.
}

\lref\XY{ X.~Yin, ``Matrix models, integrable structures, and
T-duality of type 0 string theory,'' arXiv:hep-th/0312236.
}

\lref\CKR{
B.~Craps, D.~Kutasov and G.~Rajesh,
``String propagation in the presence of cosmological singularities,''
JHEP {\bf 0206}, 053 (2002)
[arXiv:hep-th/0205101].
}

\lref\JT{
N.~Toumbas and J.~Troost,
``A time-dependent brane in a cosmological background,''
arXiv:hep-th/0410007.
}

\lref\SW{
A.~M.~Sengupta and S.~R.~Wadia,
``Excitations And Interactions In D = 1 String Theory,''
Int.\ J.\ Mod.\ Phys.\ A {\bf 6}, 1961 (1991).
}

\lref\GK{
D.~J.~Gross and I.~R.~Klebanov,
``Fermionic String Field Theory Of C = 1 Two-Dimensional Quantum Gravity,''
Nucl.\ Phys.\ B {\bf 352}, 671 (1991).
}

\lref\DV{ R.~Dijkgraaf and C.~Vafa, ``N = 1 supersymmetry,
deconstruction, and bosonic gauge theories,''
arXiv:hep-th/0302011.
}

\lref\GJM{D.~Ghoshal, D.~P.~Jatkar and S.~Mukhi, ``Kleinian
singularities and the ground ring of C=1 string theory,'' Nucl.\
Phys.\ B {\bf 395}, 144 (1993) [arXiv:hep-th/9206080].
}

\lref\GR{ D.~Gaiotto and L.~Rastelli, ``A paradigm of open/closed
duality: Liouville D-branes and the Kontsevich model,''
arXiv:hep-th/0312196.
}

\lref\Ba{ J.~L.~F.~Barbon, ``Perturbing the ground ring of $2D$
string theory,'' Int.\ J.\ Mod.\ Phys.\ A {\bf 7}, 7579 (1992)
[arXiv:hep-th/9201064].
}

\lref\KaC{ S.~Kachru, ``Quantum rings and recursion relations in
$2D$ quantum gravity,'' Mod.\ Phys.\ Lett.\ A {\bf 7}, 1419 (1992)
[arXiv:hep-th/9201072].
}

\lref\He{ S.~Hellerman, ``On the landscape of superstring theory
in D $>$ 10,'' arXiv:hep-th/0405041;
S.~Hellerman and X.~Liu, ``Dynamical dimension change in
supercritical string theory,'' arXiv:hep-th/0409071.
}

\lref\Lee{ J.~Lee, ``Time dependent backgrounds in 2-d string
theory and the S matrix generating functional,'' Phys.\ Rev.\ D
{\bf 49}, 2957 (1994) [arXiv:hep-th/9310190].
}

\lref\Vij{ V.~Balasubramanian, E.~Keski-Vakkuri, P.~Kraus and
A.~Naqvi, ``String scattering from decaying branes,''
arXiv:hep-th/0404039.
}

\lref\Oku{ K.~Okuyama, ``Comments on half S-branes,'' JHEP {\bf
0309}, 053 (2003) [arXiv:hep-th/0308172].
}

\baselineskip 18pt plus 2pt minus 2pt

\Title{\vbox{\baselineskip12pt
\hbox{hep-th/0411019}\hbox{HUTP-04/A041}
  }}
{\vbox{\centerline{Matrix Model and Time-like Linear Dilaton
Matter}}} \centerline{Tadashi Takayanagi\foot{e-mail:
takayana@bose.harvard.edu}}

\medskip\centerline{ \it Jefferson Physical Laboratory}
\centerline{\it Harvard University}
\centerline{\it Cambridge, MA 02138, USA}

\vskip .5in \centerline{\bf Abstract}
 We consider a matrix model description
 of the 2d string theory whose matter part is given by
 a time-like linear dilaton CFT. This is
 equivalent to the $c=1$ matrix model with a deformed, but
 very simple fermi surface.
Indeed, after a Lorentz transformation, the corresponding 2d
spacetime is a conventional linear dilaton background with a
time-dependent tachyon field. We show that the tree level
scattering amplitudes in the matrix model perfectly agree with
those computed in the world-sheet theory. The classical
trajectories of fermions correspond to the decaying D-branes in
the time-like linear dilaton CFT. We also discuss the ground ring
structure. Furthermore, we study the properties of the time-like
Liouville theory by applying this matrix model description. We
find that its ground ring structure is very similar to that of the
minimal string.

\noblackbox

\Date{}

\writetoc

\newsec{Introduction}

It is well-known that the two dimensional string theory with a
static linear dilaton and Liouville potential  can be described
non-perturbatively by the dual matrix model \tdsa \tdsb \tdsc,
called $c=1$ matrix model\foot{For reviews see e.g. \KlR \GM \PoR
\Nak \MaR.}.
 At the world-sheet level, this
model is equivalent to a free boson theory (with the central charge
$c=1$ matter) plus the Liouville theory ($c=25$), defined by the
world-sheet action\foot{In this paper we set $\al=1$.} and the
string coupling constant
 \eqn\usual{S=\int
d\sigma^2[-\de X^0\bar{\de}X^0+ \de\phi\bar{\de}\phi+ \mu
e^{2\phi}],\ \ \ g_s=e^{2\phi}.} There is only one propagating
scalar field $\eta$, which is related to the tachyon field $T$ in
bosonic string via $T\sim g_s\cdot \eta$. It behaves like a
massless scalar field in the 2d linear dilaton background. The
dual $c=1$ matrix model is defined by a quantum mechanics of a
$N\times N$ Hermitian matrix $\Phi$ with a inverse harmonic
potential (after the double scaling limit $N\to \infty$)
\eqn\mattt{S_{mat}=\int dt \Tr\left[(D_{t}\Phi)^2+\Phi^2\right].}
Here, $D_{t}=\de_{t}-i[A_{t},]$ denotes the covariant derivative
with respects to the $U(N)$ gauge symmetry, projecting out
non-singlet sectors. The eigenvalues $x$ of $\Phi$ behave like $N$
free fermions and they form a fermi sea. The static vacuum \usual\
of string theory corresponds to the static fermi surface
\eqn\fermis{ p^2-x^2=-2\mu,} in the two dimensional semiclassical
phase space $(x,p)\equiv (x,\dot{x})$. We can also employ the type
0 model \TT \six\ or type II model \GTT \TS\ to make the
non-perturbative issues clearer.

As a next step, it will also be natural and interesting to ask
what will happen if we consider a spacetime with a different
property in the time direction. One of the simplest examples will
be the time-like linear dilaton theory and this is a basic example
of time-dependent backgrounds in string theory\foot{Refer to
e.g.\CKR \DM \ST \He \HT \JT\ for recent discussions.}. In our
context, we can consider a string model defined by
 the time-like linear dilaton theory (with the central charge
$c=1-6q^2$) plus the space-like Liouville theory $(c=1+6Q^2)$ on
the world-sheet. This may be called a non-minimal $c<1$
non-critical string. Its world-sheet action is simply given by
\eqn\tldaction{ S=\int d\sigma^2[-\de X^0\bar{\de}X^0+
\de\phi\bar{\de}\phi+ \mu e^{2b\phi}],} with the background charge
terms which correspond to the coupling constant
\eqn\gstld{g_s=e^{qX_0+Q\phi}.} The values of the background
charges are \eqn\qq{Q=b+\f{1}{b},\ \ q=-b+\f{1}{b},} in terms of
the parameter $b$, which satisfies the condition\foot{Here, the
condition $b<1$ comes from the Seiberg bound \Se\ and also we
assumed that $b$ is positive using the sign flip $\phi\to -\phi$.}
\eqn\paramb{0<b<1.} In this rather simple example we can solve the
theory exactly by applying known results of the Liouville theory
\DO \COL \COR. It is obvious that the system will get strongly
coupled in the late time. However, if we consider the physical
process of scattering of closed strings from the Liouville wall,
the process itself does not occur in the strongly coupled region
because of the inequality $q<Q$.

After the Lorentz transformation,
\eqn\lor{\ti{X^0}=\f{Q}{2}X^0+\f{q}{2}\phi,\ \ \ \
\ti{\phi}=\f{q}{2}X^0+\f{Q}{2}\phi,} we can equivalently obtain
the usual static linear dilaton vacuum perturbed by a
time-dependent Liouville potential defined by \eqn\tdaction{
S=\int d\sigma^2\left[-\de \ti{X^0}\bar{\de}\ti{X^0}+
\de\ti{\phi}\bar{\de}\ti{\phi}+ \mu
\exp\left((b^2-1)\ti{X^0}+(1+b^2)\ti{\phi}\right)\right],\ \ \
g_s=e^{2\ti{\phi}}.} In general, time-dependent backgrounds in 2d
string theory correspond to deformed and time-dependent fermi
surfaces in the $c=1$ matrix model and this issue
 has been discussed in the
papers \PoS \MPY  \mopl \Lee \AKKT \KSone \KStwo \DDLM. As
recently pointed out in \KSone, they lead to non-perturbatively
tractable examples of the interesting time-dependent model of
closed string tachyon condensation. In this paper we would like to
closely understand the duality between the time-dependent
backgrounds in 2d string theory and the matrix model with a
deformed fermi surfaces via the special example \tdaction, where
we can solve the theory in both sides.

It is also intriguing to consider the case where $b$ is imaginary
(or $b^2<0$). This corresponds to the time-like Liouville theory
\DM \ST \SC\ after the double wick rotation $(X^0,\phi)\to
(-i\phi,-iX^0)$ in \tldaction. Since this conformal field theory
is far from well-understood, the matrix model formulation should
be definitely useful. As we will see later, indeed we find rather
different properties compared with those in the usual space-like
Liouville theory.

This paper is organized as follows. In section 2 we first give a
direct matrix model dual of the 2d string theory with the
time-like linear dilaton matter; and then we show that the model
is equivalent to the ordinary $c=1$ matrix model via a field
redefinition as expected from the Lorentz invariance. We also
compute the closed string emission from the decaying D-branes and
identify the leg factor from the results. In section 3 we give an
equivalent description as a time-dependent background in $c=1$
matrix model. We also compute the scattering S-matrices in this
background and find agreements with those in the world-sheet
theory. In section 4 we discuss the time-like Liouville theory by
applying the matrix model dual. We correctly reproduce the
expected spacetime geometry using the collective field
description. In section 5 we consider the ground ring structure of
our background and discuss the relation to
 non-compact Calabi-Yau manifolds.
In section 6 we summarize the results and discuss future problems.

\newsec{Matrix Model and 2D String with Time-like Linear
Dilaton Matter}

First, let us try to derive directly the matrix model dual of the
2d string with the time-like linear dilaton matter defined by
\tldaction\ and \gstld. To construct a matrix model for a new
background it is helpful to remember the recent interpretation of
the $c=1$ matrix model as a theory of unstable D0-branes (so
called ZZ-brane \ZZ) \MV \KMS. The matrix $\Phi$ can be regarded
as a open-string tachyon field on them and the matrix model itself
corresponds to an effective action of such D-branes. Then we can
  argue that a matrix model dual of time-like linear
dilaton background \tldaction\ is defined by \eqn\tldmat{
S_{mat'}=\int dt\ e^{-qt}\Tr\left[(D_{t}\Phi)^2+\Phi^2\right].} We
have put the time-dependent factor $e^{-qt}$ because the D-brane
action is proportional to $g^{-1}_s\propto e^{-qt}$ under the
identification $X^0=t$. We chose the tachyonic mass term in
\tldmat\ such that it agrees with the mass of the D0-brane \ZZ\
calculated in the boundary Liouville theory.

By using the gauge symmetry we can again diagonalize the matrix
into the eigenvalues $\lambda_i$. Then the action becomes
\eqn\actei{S_{mat'}=\int dt\
e^{-qt}\sum_{i}\left[\dot{\lambda}_i(t)^2+\lambda_i(t)^2\right].}
The classical trajectories in this system \actei\ are given by
\eqn\ctraj{\lambda(t)=C_1 e^{-bt}+C_2 e^{\f{1}{b}t},} where
$C_{1}$ and $C_{2}$ are arbitrary constants. They correspond to
the time-dependent open string tachyon field (so called Rolling
tachyon \Sen) on unstable D0-branes.

Actually, after the redefinition of the variable \eqn\change{
\lambda_i(t)=e^{\f{q}{2}t}x_i(t),} the action can be written as
(up to total derivative terms)
 \eqn\remat{S_{mat'}=\int dt\
\sum_{i}\left[\dot{x_i}(t)^2+(1+\f{q^2}{4})x_i(t)^2\right].} Now
we have the conventional $c=1$ matrix model with a shifted tachyon
mass. This is expected since we know that the 2d string with the
time-like linear dilaton matter is equivalent to the conventional
$c=1$ string via the Lorentz transformation \lor. Indeed if we
perform the Lorentz transformation $\f{\de}{\de
t}=\f{Q}{2}\f{\de}{\de \ti{t}} +\f{q}{2}\f{\de}{\de \ti{\phi}}$
into the usual two dimensional string theory with the linear
dilaton \tdaction , then we can derive the ordinary action of $c=1$
matrix model with the correct tachyon mass
\eqn\recmat{S_{mat}=\int d\ti{t}\
\sum_{i}\left[\dot{x_i}(\ti{t})^2+x_i(\ti{t})^2\right].}

\subsec{Closed String Emission and Leg Factor}

{}From the world-sheet viewpoint, a particular class of open
string tachyon condensation on the unstable D0-brane can be
represented by the boundary time-like Liouville theory \STL \LNT
\GS \Oku \Vij\ (so called the half S-brane) defined by the action
\eqn\blt{S=\int d\sigma^2(-\de X^0\bar{\de}X^0)+ \mu_B
\int_{\de\Sigma} d\sigma e^{-bX^0},} corresponding to the first
term in \ctraj. The second term in \ctraj\ is explained as the
dual boundary cosmological constant. This theory can be regarded
as a time-like continuation \GS \SCT\ of the boundary conformal
field theory on a FZZT-brane \FZZ \T. By using this observation,
we can compute the closed string one-point function on the
decaying D0-brane as follows\foot{We assume $\al=1$ and define the
momentum $P$ and energy $E$ such that $p^\mu=(E,P)$ and
$p_\mu=(E,-P)$. The vertex operator is given by
$e^{(q-iE)X^0+(Q+iP)\phi}$. When $P>0$ the particle is moving
toward strongly coupled region $\phi\to \infty$. Also, in the
computation of correlators, we are using a slightly different
normalization of $\mu$ (by the factor $\pi$) compared with the
paper \FZZ.} \eqn\emission{\eqalign{ &\la e^{(q-iE)X^0+(Q+iP)\phi}
\lb_{E=P} =e^{-i\f{E}{b}\log\mu_B}\cdot(\mu \gamma(b^2)
)^{-i\f{P}{2b}}\cdot\f{\Gamma(iP/b)} {\Gamma(-iP/b)}\cr & \la
e^{(q-iE)X^0+(Q+iP)\phi} \lb_{E=-P} =
e^{-i\f{E}{b}\log\mu_B}\cdot(\mu\gamma(b^2))^{-i\f{P}{2b}}
\cdot\f{\Gamma(iPb)} {\Gamma(-iPb)},}} where we have defined
$\gamma(b^2)=\f{\Gamma(b^2)}{\Gamma(1-b^2)}$; the on-shell
conditions are given by $E=P$ and $E=-P$ in the above two cases,
respectively.

The physical meaning of this one-point function is the closed
string emission from the decaying D-branes \LLM. In 2d string
theory, the closed string field is equivalent to the fluctuation
of the fermi surface via the bosonization up to the momentum
dependent phase factor called the leg-factor. On the other hand,
each fermion itself can be regarded as a decaying D0-brane \MV
\KMS. As pointed out in \KMS, we can directly confirm these
identifications from the fact that the closed string emission
amplitude is given by a phase factor which coincides with the
leg-factor (except the energy dependent term due to the
time-delay). Interestingly, we can also find a similar story in
our generalized backgrounds \tldaction \gstld. Indeed the closed
string emission \emission\ is given by a pure phase factor.
Furthermore, we can check that it is the same as the leg-factor.
To see this, consider the two point function (or reflection
coefficient) \DO \COL \COR \eqn\twop{S(P)\equiv\la
e^{(q-iP)X^0+(Q+iP)\phi}\ e^{(q+iP)X^0+(Q+iP)\phi}\lb
=-(\mu\gamma(b^2))^{-iP/b} \f{\Gamma(iP/b)
\Gamma(ibP)}{\Gamma(-iP/b)\Gamma(-ibP)}.} This is exactly the
multiplication of the two terms in \emission.

These discussions on the closed string emission and leg-factor can
be made clearer by performing the Lorentz transformation \lor\ of
these quantities into the system \tdaction. The transformed energy
and momentum are given\foot{In particular, for a massless particle
with $E=P$ we have the relation $\ti{E}=\ti{P} =E/b=P/b$, while in
the opposite case $E=-P$ we get $\ti{E}=-\ti{P} =bE=-bP$.}
 by \eqn\momel{\ti{E}=\f{Q}{2}E+\f{q}{2}P,\ \ \ \
\ti{P}=\f{q}{2}E+\f{Q}{2}P.} Then we find the closed emission from
the half s-brane \eqn\emissionl{\eqalign{ &\la
e^{-i\ti{E}\ti{X}^0+(2+i\ti{P})\ti{\phi}} \lb_{\ti{E}=\ti{P}}
=e^{-i\ti{E}\log\mu_B}
(\mu\gamma(b^2))^{-i\f{\ti{P}}{2}}\f{\Gamma(i\ti{P})}
{\Gamma(-i\ti{P})} \equiv e^{-i\ti{E}\log\mu_B}\cdot
e^{i\vp_+(\ti{P})}, \cr & \la
e^{-i\ti{E}\ti{X^0}+(2+i\ti{P})\ti{\phi}} \lb_{\ti{E}=-\ti{P}} =
e^{-i\f{\ti{E}}{b^2}\log\mu_B}
(\mu\gamma(b^2))^{-i\f{\ti{P}}{2b^2}}
\f{\Gamma(i\ti{P})} {\Gamma(-i\ti{P})}\equiv
e^{-i\f{\ti{E}}{b^2}\log\mu_B}\cdot e^{i\vp_-(\ti{P})}.}} The
phase factor $e^{i\vp_\pm(\ti{P})}$
 should be regarded as the
 leg factor in our time-dependent background of 2d string.
 Notice that $\vp_{+}$ is the same as the usual leg-factor in
 $c=1$ matrix model as is expected since we can write the
 Liouville term in \tdaction\ as $\mu \exp(2\ti{\phi})$ when
 $\ti{X^0}=-\ti{\phi}$.
When we consider a incoming wave with the energy $\ti{E}(=\ti{P})$
and its reflection, the energy of the outgoing wave is shifted
into $\ti{E}'=b^2\ti{E}(=-\ti{P}')$ due to the Doppler shift since
the the Liouville wall \tdaction\ is moving. Then the two point
function is given by \eqn\twop{S(\ti{P})\equiv\la
e^{-i\ti{P}\ti{X}^0+(2+i\ti{P})\ti{\phi}}\
e^{ib^2\ti{P}\ti{X}^0+(2+ib^2\ti{P})\ti{\phi}}\lb
=-(\mu\gamma(b^2))^{-i\ti{P}}
\f{\Gamma(i\ti{P})\Gamma(ib^2\ti{P})}
{\Gamma(-i\ti{P})\Gamma(-ib^2\ti{P})}.}
This reflection amplitude can be nicely rewritten in terms of the
leg factors \eqn\twoppp{S(\ti{P})=-e^{i\vp_+(\ti{P})}\cdot
e^{i\vp_-(b^2\ti{P})},} as expected. In this way we have confirmed
the identification of matrix model fermions with decaying
D-branes\foot{If we consider the static D0-brane (i.e. ZZ-brane
\ZZ\ in \tldaction), then naively we will obtain a moving D0-brane
at the velocity $q/Q<1$ in \tdaction\ after the Lorentz
transformation. This D0-brane
 may not be static since there is the time-dependent Liouville
 potential.
 In this matrix model, obviously
this configuration corresponds to a single fermion on
the top of the inverse harmonic potential.}
 in our time-dependent backgrounds
\tdaction.
 These results of the leg-factor will also be useful
later when we compare the scattering amplitudes in the matrix
model with the world-sheet computation for arbitrary values of
$b$.

Even though we have examined the special case $C_1\neq 0$ and
$C_2=0$ in \ctraj (i.e. half S-brane), it is natural to expect the
similar computations can be done for more general $C_1$ and $C_2$
(so called full S-brane) as has been done for $b=1$ case \KMS\ by
using the rolling tachyon boundary state \Sen. Thus our matrix
model here predicts the existence of boundary states for general
profiles of \ctraj\ in the time-like linear dilaton theory and its
construction will be an intriguing future problem. Since the
trajectory corresponding to the D-brane should be above the fermi
level in the matrix model, we can find a bound
$|C_1|^{b^2}|C_2|\leq\mu$, where $\mu$ is the fermi level for our
background and will be defined in the next section.

\newsec{Equivalent Time-dependent Background in $c=1$ Matrix Model}

As we have seen in the previous section, the matrix model dual of
the 2d string background \tldaction\ can be given by a
time-dependent background of $c=1$ matrix model. The 2d Lorentz
transformation is not clear in the holographic dual matrix model
since the Liouville direction is hidden inside the infinitely many
eigenvalues. Thus it is an non-trivial and intriguing problem to
realize the time-dependent background \tdaction\ in the $c=1$
matrix model\foot{In most of the literature (e.g. \AKKT \KSone
\KStwo \DDLM), a non-zero static cosmological constant as in
\usual\ is assumed to compute physical quantities. In particular,
it is possible to solve the matrix model for an Euclidean
compactified time by applying the Toda Lattice integrable
structure \DMP \EK \KKK \Ko \XY\ for a rather general backgrounds
with time-dependent tachyon perturbations as shown in \AKKT . In
the discussions of the present paper, however, we do not put the
static cosmological constant term $\sim e^{2\phi}$ (for $b\neq 1$)
because it will change the asymptotic behavior and lead to a
different theory. Interestingly, this suggests that our
backgrounds may be related to a new integrable structure of $c=1$
matrix model.}.

We argue that the string theory background \tdaction\ can be
identified with the time dependent fermi surface  in $c=1$ matrix
model (we assume $\mu>0$) \eqn\tdfer{
(-p-x)^{b^2}(p-x)=2^{1+b^2}\mu~ e^{(b^2-1)\ti{t}},} where
$\ti{t}(=\ti{X^0})$ is the time\foot{ Notice the relation
$\ti{t}\sim \f{Qt}{2}$ due to the Lorentz transformation.} in the
matrix model (see \recmat). Its qualitative behavior can be
summarized as follows. Because of the condition $b^2<1$, in the
far past $t\to -\infty$, the fermi surface is pushed into the
infinity and there is no fermi sea. After that, the fermi sea
gradually begins to appear from the weakly coupled region
$|x|>>1$, and it finally spreads out completely. This is
intuitively consistent with the property of the time-dependent
tachyon field in \tdaction. We starts with the infinite tachyon
condensation, which means that spacetime disappears. Then the
tachyon field becomes smaller and the spacetime appears.
Eventually, the tachyon field becomes zero and we have the
ordinary (strongly coupled) 2d spacetime with the linear dilaton.

In order to see that \tdfer\ is consistent with the time
evolution, we can rewrite it simply as follows
\eqn\costt{W_{1,0}^{b^2}\ W_{0,1}=2^{1+b^2}\mu,} by using the
conserved quantities (or the classical $w_\infty$ generators \AJ
\PoR ) for each fermion \eqn\winft{W_{1,0}=-(p+x)e^{-\ti{t}} ,\ \
\ W_{0,1}=(p-x)e^{\ti{t}}.}

Also since we are discussing the 2d bosonic string, only one fermi
surface is relevant and we can only consider the fermi surface
which satisfies the constraints $p+x<0$ and $p-x>0$. The matrix
model defined by the two fermi surfaces, replacing $-p-x$ with
$|p+x|$ in \tdfer\ describes a background with a time-dependent
NSNS scalar field in type 0B string theory. Though we mainly
restrict to the bosonic string case below, we can obtain the
almost same result in the type 0 case as we will briefly comment
later.

The profile of the semiclassical fermi surface includes all
information on the dual string theory at tree level, and can be
uniquely determined from the action \tdaction\ as we will see
later\foot{To be exact, we should say that the cosmological
constant $\mu$ in \tdfer\ corresponds to $\mu\gamma(b^2)$ in
\tldaction\ (see appendix A).}. To go beyond the tree level we
need to define a time-dependent quantum state in the matrix
quantum mechanics \mattt\ and this problem is beyond the scope of
this paper.

In general, when a fermi surface is given, the expectation value
of tachyon field in the asymptotic region $\phi\to -\infty$ can be
determined by its deviation from the singular fermi surface
$p^2-x^2=0$. We can write this in the following way, \eqn\polff{
p_{\pm}\simeq\mp x\pm \f{\ep_{\pm}}{x}\ \ \ (x\to-\infty),} where
$p_{+}$ (or $p_{-}$) is the value of the momentum at the upper (or
lower) branch of the fermi surface \tdfer. After identifying the
spacial coordinate as $x=e^{-\ti{\phi}}$, the deviations
$\ep_{\pm}$ are related to the left and right-moving part of the
massless scalar field $\eta$ in the 2d spacetime \PoS\ via
\eqn\fermisd{\eqalign{(\de_{\ti{t}}-\de_{\ti{\phi}})\
\eta(\ti{t},\ti{\phi}) &=\pi^{-1/2}\ep_+(\ti{t}-\ti{\phi}), \cr
(\de_{t}+\de_{\phi})\
\eta(\ti{t},\ti{\phi})&=-\pi^{-1/2}\ep_-(\ti{t}+\ti{\phi}).}} This
is explained by the bosonization of Dirac fermions and the
massless scalar field $\eta$ is the collective field of the fermi
sea \DJ \PoS \SW \GK \PoR.  The scalar field $\eta$ is related to
the tachyon field $T$ in 2d bosonic string as follows
\eqn\relationco{ T(\ti{t},\ti{\phi})=e^{2\ti{\phi}}\cdot
\eta(\ti{t},\ti{\phi}),} up to the leg factor.

Let us apply this method to \tdfer\ in order to examine the
tachyon field in this background. If we expand the fermi surface
near the two asymptotic regions as in \fermisd, we get \eqn\asyme{
p-x\sim 2\mu e^{(b^2-1)\ti{t}}|x|^{-b^2},\ \ \ p+x\sim-
2\mu^{1/b^2} e^{(1-1/b^2)\ti{t}}|x|^{-1/b^2}.} Following the rule
\fermisd, we can extract the expectation value of the tachyon
field from \asyme\
\eqn\asyta{T_{-}=\mu\exp\left((b^2-1)\ti{X^0}+(1+b^2)
\ti{\phi}\right),\ \ \
T_{+}=\mu^{1/b^2}\exp\left((1-1/b^2)\ti{X^0}+(1+1/b^2)
\ti{\phi}\right).} The two tachyon fields $T_{-}$ and $T_{+}$
represent the two contributions from each term in \asyme. The
first one $T_{-}$ exactly coincides with the Liouville potential
in \tdaction. The second one also agrees with the Lorentz
transformation of the dual Liouville potential
$\ti{\mu}e^{\f{2}{b}\phi}\ \ \ (\ti{\mu}=\mu^\f{1}{b^2})$. As is
known in the Liouville conformal field theory, the dual potential
automatically appears whenever we put the original one \DO \COL
\COR \MaR. Thus our matrix model description precisely reproduces
this fact. In terms of the dual picture we can also rewrite
\tdfer\ as follows \eqn\tdferd{ (-p-x)(p-x)^{\f{1}{b^2}}=\ti{\mu}
e^{(1-1/b^2)\ti{t}}.} Notice that the form of the fermi surface
\tdfer\ is determined uniquely by the identification of asymptotic
fields and the time-evolution.

{}From the viewpoint of the matrix model \tldmat, which is
directly dual to the background \tldaction\ before the Lorentz
transformation, the fermi surface is given by
\eqn\fermist{\left(-\dot{\lambda}-b\lambda\right)^{b^2}
\left(\dot{\lambda}-\f{1}{b}\lambda\right)=2^{1+b^2}\mu.} This
looks like a static background and is consistent with the static
Liouville potential in \tldaction. To find the result \fermist ,
notice that the conserved quantities are now given by
\eqn\winft{W_{1,0}=-(\dot{\lambda}+b\lambda)e^{-t/b},\ \ \
W_{0,1}=(\dot{\lambda}-\lambda/b)e^{bt}.}

\subsec{Scattering Amplitudes}

To find a further evidence that the fermi surface \tdfer\ is dual
to the background \tldaction\ or equally \tdaction, it is useful
to compare the scattering S-matrices. To compute the scattering
amplitudes in the matrix model side, we can apply the Polchinski's
scattering equation \PoS\
\eqn\scatter{\epsilon_{+}(\ti{t}-\ti{\phi})=\ep_{-}(\ti{t}
-\ti{\phi}-\log
(\ep_{+}(\ti{t}-\ti{\phi})/2)),} where $\ep_{+}$ and $\ep_{-}$ are
the incoming and outgoing deformations of the fermi surface
defined previously in \polff. This equation states that an incoming
wave completely turns into the outgoing one by the reflection with
a time-delay represented by the
 $-\log\ep_{+}$ term in \scatter.

We can express excitations from \asyme
\eqn\asymmp{\ep_+=2\mu^{1/b^2}
e^{(1-1/b^2)(\ti{t}-\ti{\phi})}(1+\delta_{+}(\ti{t}-\ti{\phi})),\
\ \ep_-=2\mu
e^{(b^2-1)(\ti{t}+\ti{\phi})}(1+\delta_{-}(\ti{t}+\ti{\phi})).} To
make the expression simple, we
 can introduce
\eqn\defs{\ti{\delta}_{-}(x)=\delta_{-}\left(\f{x}{b^2}
-\f{1}{b^2}\log\mu
\right).} Then the scattering equation \scatter\ becomes
\eqn\solsca{(1+\delta_{+}(x))^{b^2}=1+\ti{\delta}_{-}
\left(x-b^2\log(1+\delta_{+}(x))\right).} We can solve \solsca\
recursively up to the order $O(\ti{\delta}^3_{-})$,
\eqn\solscare{\eqalign{
\delta_{+}=&\f{1}{b^2}\ti{\delta}_{-}+\left(-\f{1}{b^2}
\ti{\delta}_{-}\ti{\delta}'_{-}+\f{1-b^2}{2b^4}
\ti{\delta}^2_{-}\right)
\cr
 &\ \ +\Bigl(\left(\f{1}{6b^6}-\f{1}{2b^4}+\f{1}{3b^2}\right)
 \ti{\delta}^3_{-}
+\f{1}{b^2}\ti{\delta}_{-}\ti{\delta}^{'2}_{-}
+(\f{3}{2b^2}-\f{1}{b^4})
\ti{\delta}^2_{-}\ti{\delta}'_{-}+\f{1}{2b^2}
\ti{\delta}^2_{-}\ti{\delta}''_{-}\Bigr). }} Notice that the
leading relation $\delta_{+}(x)\sim
\f{\delta_{-}(x/b^2+const.)}{b^2}$ tells us that the incoming wave
with energy $\ti{E}$ will be shifted into the energy $b^2\ti{E}$
due to the moving wall. The relation \solscare\ shows the $1\to
1,1\to 2$ and $1\to 3$ scattering\foot{The term `$n\to m$
scattering' means that the process with $n$ incoming and $m$
outgoing particles.} of closed strings. As we will show in the
last of this section, it is also possible to find the exact
solution to \solsca.

In order to compare these results with those of the string theory
scattering amplitudes in the background \tldaction , we would like
to perform the Lorentz transformation \lor. The massless scalar
field $\eta$ can be written in terms of the deformation of fermi
surface by using \fermisd\
\eqn\fermisd{\eqalign{(\de_{t}-\de_{\phi})\eta(t,\phi)&=\pi^{-1/2}
\Delta_+(t-\phi), \cr
(\de_{t}+\de_{\phi})\eta(t,\phi)&
=-\pi^{-1/2}\Delta_-(t+\phi+(\log\mu)/b),}}
where $\Delta_{\pm}$ is defined by \eqn\flucd{\eqalign{
\Delta_{+}(y)&=2b\mu^{1/b^2}e^{-qy}
\cdot\delta_{+}(by),\cr
\Delta_{-}(y)&=\f{2}{b}\mu^{1/b^2}e^{-qy}
\cdot\ti{\delta}_{-}(by).}}
 By substituting \flucd\ into \solscare\ , we
can find the scattering equation in the original frame
\eqn\scatterlo{\Delta_{+} =\Delta_{-}
-\f{1}{4}\mu^{-1/b^2}(e^{qy}\Delta_{-}^2)'+\f{b}{24}\mu^{-2/b^2}
(e^{2qy}\Delta_{-}^3)'+\f{1}{24}\mu^{-2/b^2}(e^{2qy}
\Delta_{-}^3)''.} The quantization of $\eta$ can be done as
\eqn\quantis{
\eta=\f{i}{2\pi^{1/2}}\int^{\infty}_{-\infty}\f{dE}{E} \left(a_E
e^{iE(t-\phi)}+\ti{a}_E e^{iE(t+\phi)}\right).} The creation and
annihilation operator satisfy (we follow the convention in \PoS)
\eqn\comute{
[a_{E},a_{E'}]=[\ti{a}_{E},\ti{a}_{E'}]=-E\cdot\delta(E+E').}
$a_{E}$ $(E>0)$ (or $\ti{a}_{E}$ $(E>0)$) represents a creation
operator of incoming (or outgoing) particle. Then the
$\Delta_{\pm}$ can be expressed as
\eqn\deltaa{\eqalign{\Delta_+(y)&=-\int \f{dE}{E}\ a_E\ e^{iEy},
\cr \Delta_-(y)&=\int \f{dE}{E}\ \ti{a}_E\
e^{iEy}\mu^{-\f{i}{b}E}.}} By plugging \deltaa\ in \scatterlo, we
obtain \eqn\scatteringre{ \eqalign{-\mu^{\f{i}{b}E}\cdot
a_E=&\ti{a}_E-\f{i}{4}\mu^{-1}E\int dE' \
\ti{a}_{E'}\cdot\ti{a}_{E-E'+iq}\cr &+\f{1}{24}\mu^{-2}(ibE-E^2)
\int dE'dE''\
\ti{a}_{E'}\cdot\ti{a}_{E''}\cdot\ti{a}_{E-E'-E''+2iq}.}} The
first term in the right-hand side represents the reflection
amplitude (or two point function) and is precisely the same as the
one \twop\ obtained in the world-sheet computation after we
multiply the previous leg factors in \emission \eqn\legfffff{
\f{\Gamma(iP/b)}{\Gamma(-iP/b)}\cdot\f{\Gamma(ibP)}{\Gamma(-ibP)},}
and perform a scaling
$\mu\to \mu\gamma(b^2)$. As we have explained in
section2, each $\Gamma$ function ratio in \legfffff\ comes from
the incoming or outgoing process, respectively. In this way, we
can read off S-matrices from \scatteringre\ including the leg
factor \legfffff\  \eqn\smatrixtld{\eqalign{&S^{(2)}_{1\to
1}(E_1,E_2) =-\delta(E_1+E_2)\cdot\mu^{-iE_1/b}\cdot
\f{\Gamma(iE_1/b)} {\Gamma(-iE_1/b)}
\cdot\f{\Gamma(-ibE_2)}{\Gamma(+ibE_2)}E_1, \cr &S^{(3)}_{1\to
2}(E_1,E_2,E_3) \cr &= \f{i}{2}\delta(E_1+E_2+E_3+iq)
\cdot\mu^{-1-iE_1/b}\cdot\f{\Gamma(iE_1/b)} {\Gamma(-iE_1/b)}
\f{\Gamma(-ibE_2)}{\Gamma(+ibE_2)}
\f{\Gamma(-ibE_3)}{\Gamma(+ibE_3)}E_1E_2E_3, \cr &S^{(4)}_{1\to
3}(E_1,E_2,E_3,E_4) \cr &= -\f{1}{4}\delta(E_1+E_2+E_3+E_4+2iq)
\cdot\mu^{-2-iE_1/b}\cdot\f{\Gamma(iE_1/b)} {\Gamma(-iE_1/b)}
\f{\Gamma(-ibE_2)}{\Gamma(+ibE_2)}
\f{\Gamma(-ibE_3)}{\Gamma(+ibE_3)}
\f{\Gamma(-ibE_4)}{\Gamma(+ibE_4)}\cr &\  \cdot
(ib-E_1)E_1E_2E_3E_4. }} As we show the details in appendix A, we
can see that these amplitudes from the matrix model
precisely agree with the
string theory results computed in \DK.

It is also possible to solve the scattering equation \solsca\
exactly by generalizing the method developed in \mopl. To find the
exact solution, we first consider the infinitesimal variation of
$\delta_{\pm}(x)$ and take the Fourier transformation. Then we
obtain the solution to \solsca\ \eqn\exactsol{
\delta_{+}(x)=\f{1}{b^2}\sum_{n=1}^{\infty}
\f{\Gamma(-\de_{x}+\f{1}{b^2})} {n!\cdot
\Gamma(-\de_{x}+\f{1}{b^2}+1-n)}\cdot (\ti{\delta}_{-}(x))^n.}
Plugging \flucd\ into \exactsol\ we get in the end
\eqn\solfin{\Delta_{+}(y)=\sum_{n=1}^{\infty}
\left(\f{b\mu^{-1/b^2}}{2}\right)^{n-1}\cdot
\f{\Gamma(-\f{1}{b}\de_{y}+1)} {n!\cdot
\Gamma(-\f{1}{b}\de_{y}+2-n)} \cdot \left
(e^{(n-1)qy}\Delta_{-}(y)^n\right).} It is easy to see that the
specific terms in \solfin\ of $n=1,2,3$ reproduce \scatterlo. It
is natural to believe that these agreements go over to general
$n\to m$ scattering amplitudes as was true \DK\ in the usual
vacuum \fermis\ (i.e. the spacial case $b=1$). In the appendix B
we also estimated the free energy at tree level and that also
agrees with the scaling behavior predicted from string theory. In
this way we have confirmed that the matrix model with the fermi
surface \tdfer\ reproduces the string theory S-matrices in the
background \tldaction.

\subsec{Brief Comments on Type 0 String Cases}

It is also possible to extend the above results to the 2d type 0
string \TT \six\ in order to consider a non-perturbatively
sensible theory (only in this subsection we set $\al=1/2$).

In the type 0B case, there are two copies of the fermi surface
\tdfer. We can choose the parameter $\mu$ independently for each
of the two surfaces and write them as $\mu_1$ and $\mu_2$. Then in
this background, there are a non-zero tachyon field $T$ and
RR-scalar field $C$ given by \eqn\typeo{T=(\mu_1+\mu_2)\cdot
e^{(b^2-1)\ti{X^{0}}+(1+b^2)\ti{\phi}},\ \ \ C=(\mu_1-\mu_2)\cdot
e^{(b^2-1)\ti{X^{0}}+(b^2-1)\ti{\phi}},} corresponding to the
symmetric and asymmetric part with respect to the exchange of the
two fermi sea.

If we consider the type 0A case in the RR-flux background, things
become more non-trivial. The equation \tdfer\ is no more
consistent with the time-evolution since the Hamiltonian is given
\six \Ka\ by that of the deformed matrix model \JY\
\eqn\pota{2H=p^2-x^2+\f{M}{x^2},\ \ \ M\equiv q^2-\f{1}{4},} where
the integer $q$ represents the background RR-flux. It is useful to
notice the conserved quantities\foot{Also notice the relation
$W_{+}W_{-}=4(M+H^2)$.} \eqn\winf{\eqalign{ &W_{+}=e^{-2t}\left(
(p+x)^2+\f{M}{x^2}\right), \cr
&W_{-}=e^{2t}\left((p-x)^2+\f{M}{x^2}\right).}} Then it is natural
to expect the fermi surface \tdfer\ in bosonic string is now
replaced by \eqn\zeroa{W_{+}^{b^2}W_{-}=\mu'^2,} in the 0A model.
When there is no RR-flux $M\simeq 0$, it is obvious that the
parameter $\mu'$ corresponds to that of the tachyon field (setting
$\mu'=\mu$ in \tdaction) because the equation \zeroa\ becomes the
same as \tdfer. The comparison with the string theory results go
over in the same way. In the non-zero RR-flux cases, the precise
relation between $\mu'$ and $\mu$ will be a bit complicated and
will be $q$ dependent.

\newsec{Matrix Model and Time-like Liouville Theory}

It is possible to add the time-like Liouville potential term
\eqn\tll{S_{1}=\nu \int d\sigma^2 e^{-2bX^0},\ \ or\ \
 S_{2}=\nu' \int d\sigma^2 e^{2X^0/b},} to our model
 defined by \tldaction\ and \gstld\ at least perturbatively.
 Note that we can put
the term \tll\ in addition to the conventional Liouville term
because the two CFTs , i.e. time and space-like ones are
decoupled.

The time $X^0$ part of this kind of CFT was considered in \ST \SC\
by assuming the analytical continuation from the usual space-like
Liouville theory. The model is obviously a basic example of
rolling closed string tachyon condensation. Though there are
several evidences that such a treatment gives sensible results,
their properties are far from well-understood. For example, the
two potentials in \tll\ are at least formally dual to each other
if we extend the result for usual space-like Liouville theory to
our time-like case. However, this looks rather strange since the
two tachyon fields behave oppositely.

\subsec{Matrix Model Dual of Time-like Liouville Theory}

On the other hand, if we know its matrix model dual, we can define
such a theory non-perturbatively. We can rewrite \tll\ as
deformations of fermi surface as we have done previously using
\polff\ \eqn\fluctu{\ep'_-=\nu e^{-(1+b^2)(\ti{t}+\ti{\phi})}=\nu
e^{-(1+b^2)\ti{t}}|x|^{1+b^2},\ \ \ \ep'_+=\nu'
e^{-\f{1+b^2}{b^2}(\ti{t}+\ti{\phi})}=\nu'
e^{+\f{1+b^2}{b^2}\ti{t}} |x|^{\f{1+b^2}{b^2}}.} These two
perturbations of fermi surface represent the background tachyon
fields, i.e. the Lorentz transformation of \tll\ \eqn\tac{T'_-=\nu
e^{-(1+b^2)\ti{X^0}+(1-b^2)\ti{\phi}},\ \ \ \ \
T'_{+}=\nu'e^{\f{1+b^2}{b^2}\ti{X^0}+\f{b^2-1}{b^2}\ti{\phi}}.}

We can argue that the fermi surface\foot{The special case $b=1$
has been discussed in \KSone\ from the viewpoint of closed string
tachyon condensation and cosmology.} is now given by
 \eqn\fermiso{(p-x)(-p-x)^{b^2}=\mu
e^{-(1-b^2)\ti{t}} +\nu (-p-x)^{2b^2} e^{-(1+b^2)\ti{t}},} by
considering\foot{However, this form \fermiso\ cannot be the unique
choice. For instance, we can assume another fermi surface
\eqn\uniq{ (p-x)^{1/b}(-p-x)^{b}=\mu^{1/b} e^{-(1/b-b)\ti{t}}
+\nu^{1/b} (-p-x)^{2b} e^{-(1/b+b)\ti{t}}.} This will also have
the same properties as \fermiso\ within the discussions in this
section since both have the same asymptotic behavior. To find the
unique fermi surface for the string theory \tdaction, we need to
compare physical quantities explicitly as we have done in the
previous section. In this paper we will not go into that detail.}
a suitable deformation of \tdferd. We assume $p+x<0,\ p-x>0$ and
$0<b<1$, and consider only bosonic string case, though the
generalization to type 0 case is possible as in section 3.2.
Indeed, the asymptotic tachyon field\foot{Here we omit the
detailed coefficients in front of $\mu$ and $\nu$.} for \fermiso\
found from \polff\ is given by the sum of $T_{\pm}$ and $T'_{-}$.
It also deserves our attention that when $\mu=0$ we can exactly
regard \fermiso\ as the analytical continuation $b\to ib$ of
\tdfer.

The time evolution of this fermi sea can be summarized as follows.
At an early time $t\to -\infty$, the fermi sea is
completely
 pushed into the infinity and thus there is no spacetime. Then
  the fermi sea begins to appear as the closed string tachyon
 field $T'_{-}$ becomes smaller. Finally for a large positive $t$,
 the fermi surface approaches the previous one \tdfer\ and
 eventually at $t=\infty$ the spacetime looks like a linear
 dilaton background. Notice that this shows that the other
 tachyon field $T'_{+}$ is not
relevant for this matrix model background.

\subsec{Spacetime Geometry from Matrix Model}

Next we want to check if the 2d spacetime obtained from \fermiso\
is indeed the same as what we expect from the string theory side.
This is much more non-trivial than the previous case \tdfer\ since
the asymptotic behavior at the early time is rather different from
the canonical one $p=\pm x$  due to the second term. To see this
it is helpful to derive the corresponding collective field theory
\DJ\ and try to find how the spacetime looks like. Intuitively,
the infinitely long spacial direction of the 2d spacetime is
dynamically generated from the infinitely extended fermi surface.
Fluctuations on the fermi surface correspond to the collective
excitations of the fermions and this is conveniently described by
the collective field theory \DJ. The collective field $\vp$ is
originally defined by the density of eigenvalues
\eqn\coll{\vp(x,\ti{t})=\Tr \delta(x-\Phi(\ti{t})).} A fluctuation
from its classical value $\vp_0=\f{1}{2\pi}(p_{+}-p_{-})$
corresponds to a massless scalar field $\eta$ (or tachyon field
$T$ in bosonic string via $T=g_s\cdot \eta$). Therefore one way to
know the properties of the spacetime is to investigate
propagations of fluctuations on the fermi surface. As pointed out
in \Al\ (see also \KStwo \DDLM), we can extract an effective
geometry of the spacetime by computing the kinetic term of $\eta$
at the quadratic order in the collective field theory, given by
\eqn\kine{S_{(2)}=\int dt\f{dx}{p_+-p_-}
[(\de_{\ti{t}}\eta)^2+(p_+
+p_-)\de_{\ti{t}}\eta\de_{x}\eta+p_+p_-(\de_{x}\eta)^2].} Here
again $p_+$ and $p_-$ denote the upper and lower branches of fermi
surface in the $(x,p)$ plane. The `effective metric' can be found
by just comparing \kine\ with the standard expression $\sim
\s{g}g^{\mu\nu}\de_{\mu}\eta\de_{\nu}\eta$ up to the conformal
transformation\foot{This means that we can always find a
coordinate where the metric is flat as noted in \Al. Here we use
the effective metric to see if the coordinate we assumed is
singular in that region.}.

Let us apply this method to our example. We can conveniently
choose the spacial coordinate $\sigma$ as follows
\eqn\paratt{-p-x=\mu^{\f{1}{1+b^2}}e^{\sigma},\ \ \ \
p-x=\mu^{\f{1}{1+b^2}}e^{-b^2\sigma+(b^2-1)\ti{t}}+\nu
\mu^{\f{1}{1+b^2}}e^{b^2\sigma-(1+b^2)\ti{t}}.} We can find two
solutions of $p$ to \fermiso\ for fixed $x$. We parameterize the
two branches by $p_{+}=p(\sigma,\ti{t})$ and
$p_{-}=p(\ti{\sigma},\ti{t})$ by introducing another function
$\ti{\sigma}(\sigma,\ti{t})$ such that $x(\sigma)=x(\ti{\sigma})$.
The parameters take the values \eqn\val{-\infty<\sigma\leq
\sigma_0(\ti{t}),\ \ \ \sigma_0(\ti{t})\leq \ti{\sigma} <\infty,}
where $\sigma_0$ is a time-dependent function and behaves like
$\sigma_0(\ti{t})\sim -\f{1+b^2}{1-b^2}\ti{t}$ for large $\ti{t}$.

Now, we can rewrite the effective field theory \kine\ in terms of
the coordinate $(\ti{t},\ti{\sigma})$. Let us consider the
asymptotic geometry, i.e. we assume that $|\ti{t}|$ and
$\ti{\sigma}$ are large, to make the computations simple. When the
two conditions (the first one just corresponds to \val)
\eqn\boundf{(1-b^2)\ti{\sigma}+(1+b^2)\ti{t}>0,\ \ \
(1+b^2)\ti{\sigma}+(1-b^2)\ti{t}>0,} are satisfied, the first and
second exponential terms (i.e. the $\nu$ independent ones) in the
right-hand side of
\paratt\ are dominant for the large values of $|\ti{t}|$ and
$\ti{\sigma}$. In this case, $(\ti{t},\ti{\sigma})$ coincides with
the coordinate $(\ti{X^0},-\ti{\phi})$ in the string theory side.
Indeed, the kinetic term of \kine\ takes the standard form $\sim
(\de_{\ti{t}}\eta)^2-(\de_{\ti{\sigma}}\eta)^2$. It is also
possible to see that for the other values than \boundf, the
effective `metric' obtained from \kine\ degenerates into that of a
line and thus this does not contribute to the spacetime geometry.
Thus we can conclude that the spacetime is given by the region
\boundf\ or equally \eqn\spacetime{\{(\ti{X^0},\ti{\phi})|\
(1-b^2)\ti{\phi}-(1+b^2)\ti{X^0}<0,\ \
(1+b^2)\ti{\phi}-(1-b^2)\ti{X^0}<0\}.}  Indeed, this is consistent
with the expectation in the world-sheet theory side. The two
conditions in \spacetime\ correspond to the tachyon walls
 $T'_{-}$ in \tac\ and $T_{\pm}$ in \asyta, respectively.
 In other words, if we return to the the original frame, the
 condition
 \spacetime\
just means the upper bound for $X^0$ and the lower bound for
$\phi$. It is again confirmed that the other tachyon field
$T'_{+}$ does not contribute in this background. This will be a
good lesson when we analyze the time-like Liouville CFT.

Then one may ask what is the matrix model configuration dual to
the tachyon field $T'_{+}$. If we remember the dual equivalent
expression of the fermi surface \tdferd, we can easily identify it
with \eqn\fermist{(-p-x)(p-x)^{1/b^2}=\mu^{1/b^2}
e^{-(1/b^2-1)\ti{t}} +\nu' (p-x)^{2/b^2} e^{(1+1/b^2)\ti{t}}.} The
previous arguments can also be applied to this case similarly. If
we simply assume $\mu=0$, then the two different backgrounds
defined by $T'_{-}$ and $T'_{+}$ correspond to the upper and lower
region divided by the surface $p-x=\nu
e^{-(1+b^2)\ti{t}}|p+x|^{b^2}$, respectively.

\newsec{Ground Ring and Possible Relations
to Non-Compact Calabi-Yau}

So far we have investigated the equivalence between the 2d string
theory in our specific backgrounds and its dual matrix model
description by looking at the properties of the tachyon field in
the 2d spacetime. There is another helpful proposal \witteng\ that
we can directly relate the fermi surface to the ring structure, so
called ground ring, of BRST invariant operators at ghost number
zero. In the ordinary static $c=1$ vacuum this is simply
given\foot{Here we mean that this relation does hold for the
on-shell tachyon states as clarified in \SH. The author thank
David Shih for explaining this point.}by \eqn\grsb{{\bf x} {\bf
y}=\mu,} as proposed in \witteng\ and proved in \six\ explicitly
(${\bf x}$ and ${\bf y}$ are the ground ring generators and will
be defined below more generally). Indeed this agrees with the
fermi surface equation \fermis\ after a rather trivial change of
basis. Let us apply this idea to our examples\foot{The author
especially thank Davide Gaiotto and Cumrun Vafa for very useful
suggestions and comments on this section.}.

When the value $b^2$ takes rational values $0<\f{p}{q}\leq 1$ ($p$
and $q$ are coprime integers), we can write the fermi surface
equation \tdfer\ in the form
\eqn\fermig{W_{0,1}^qW_{1,0}^{p}=\mu^q,} using the conserved
quantities \winft. This strongly implies that the ground ring
structure \witteng\ in our background \tldaction\ will be
\eqn\ground{{\bf x}^q {\bf y}^p=\mu^q,} where ${\bf x}={\bf
a\bar{a}}$ and ${\bf y}={\bf b\bar{b}}$ are the ground ring
generators. We can write them explicitly via the Lorentz
transformation (we show only the ones in the left-moving sector)
\eqn\grounge{{\bf a}=\Bigl (cb+\s{\f{q}{p}}\ \de(\phi+iX) \Bigr)
e^{\s{\f{p}{q}} (iX-\phi)},\ \ \ \ {\bf b}=\Bigl (cb+\s{\f{p}{q}}\
\de(\phi-iX) \Bigr) e^{-\s{\f{q}{p}}(iX+\phi)},} where $X=iX^0$ is
the Euclidean time. Obviously for $p=q=1$ this statement is
reduced to the basic result \grsb. For general $p$ and $q$, we
will be able to show this relation almost in the same way.

These expressions \grounge\ are formally the same as the ground
ring generators for the $(p,q)$ minimal string in the coulomb gas
description \KMSG \GOV. Nevertheless the ground ring structure
\ground\ for our non-minimal case is different from that of the
minimal string found in \SH\ because there are screening operators
in the minimal model case. On the other hand, if we consider
another one \fermiso\ (or \uniq) corresponding to the time-like
Liouville potential, we obtain the relation at $\mu=0$ \eqn\grt{
{\bf y}^p\cdot ({\bf x}^q-\nu^q\ {\bf y}^{p})=0.} This looks very
close to the one in the minimal string \SH \ADKMV, except the
factor ${\bf y}^p$. This may be natural since we now have the
Liouville potential (or screening operator) in the matter CFT as
in the minimal case. We would like to leave the details on this
issue for future work.

As pointed out in \six, the values of the ground ring elements are
also directly related to the charges carried by decaying
D0-branes\foot{In the papers \SenC, another definition of the
conserved charges carried by the D0-branes was considered. This
may also lead to similar interesting results in our case.}.
Generalizing this analysis to our case, we can show that the
expectation value of ${\bf x}$ and ${\bf y}$ on a decaying
D0-brane discussed in section 2.1 is given by (up to a constant)
\eqn\expect{\la {\bf x}\lb=\mu_B\mu^{\f{1}{2}},\ \ \ \la {\bf
y}\lb=\mu_B^{\f{1}{b^2}} \mu^{\f{1}{2b^2}},} employing the
boundary Liouville theoretic results \FZZ \T. The $\mu_B$
dependence of \expect\ is indeed consistent\foot{Note that here we
only discuss the `half S-brane' \STL \LNT \GS. For more general
boundary interactions like an analogue of the `cosh' brane \Sen,
we will expect a non-trivial renormalization of $\mu_B$ as is so
in the $c=1$ CFT.}
 with the expectation values
of $W_{0,1}$ and $W_{1,0}$ for the trajectory $-\lambda(t)=\mu_B
e^{-bt}+\ti{\mu}_B e^{t/b}$, where we have included the dual
cosmological constant $\ti{\mu}_B=\mu_B^{1/b^2}$. It would also be
intriguing to study the other kind of D-branes (or non-compact
branes) in these spaces and compare them with the dual 2d
dimensional string in order to understand the open-closed duality
\ADKMV \GR.

As is well known, the $c=1$ string at the self-dual radius is
equivalent to the topological string (B-model)
on the conifold \GV\ (refer
to \Ba \KaC \GJM \DV\ for more general backgrounds obtained from
the quotients or perturbations of $c=1$ string, and also refer to
\ADKMV\ for modern perspectives.). Thus we may expect that our
backgrounds, when suitably compactified, will also be dual to the
topological string on specific non-compact Calabi-Yau manifolds.
After we wick-rotate the time into the Euclidean one, we can
impose the periodicity\foot{This is a different compactification
radius than the one $R=\s{\f{p}{q}}$ in the Coulomb-gas
representation of the minimal model. This is because the latter
has the screening operators. Our model does not have such
operators and thus this smaller radius is not consistent with
$g_s$.} $X\sim X+2\pi \s{pq}$ (i.e. the radius $R=\s{pq}$ in the
$\al=1$ unit) since the coupling constant $g_s\propto e^{-iqX}$
respects this\foot{A similar compactification in a time-like
direction also recently
discussed in a matrix model dual of harmonic oscillator \IM.}. The
most plausible speculation will be that this compactified
background is dual to the non-compact Calabi-Yau manifold defined
by \eqn\groundring{ {\bf x}^q{\bf y}^p+{\bf w}{\bf z}=\mu^q.} We
can choose the corresponding ground ring generators so that the
momentum and winding number obey the standard quantization rule
\eqn\detail{ {\bf x}={\bf a\bar{a}},\ \ {\bf y}={\bf b\bar{b}},\ \
{\bf w}={\bf a}^q{\bf\bar{b}}^p,\ \ {\bf z}={\bf b}^p{\bf
\bar{a}}^q.} It will be possible to find a similar algebraic
equation for type 0 string case (see \six \INOS \TS \DOV\ for
relevant discussions on the $\hat{c}=1$ string).

It may be helpful to compare this with the known ground ring for
the $c=1$ string at the radius $R=\f{r}{s}$ ($r$ and $s$ are
coprime integers), given by \eqn\rqp{({\bf x}{\bf y})^r+({\bf
w}{\bf z})^s=\mu',} as found in \GJM . This is obviously a
different background from ours. However, it is
curious that in the special case of the common radius $R=q\in {\bf
Z}$, this equation \rqp\ agrees with ours \groundring.

\newsec{Conclusions and Discussions}

In this paper we have discussed a matrix model dual of the 2d
string theory with a time-like linear dilaton matter. This may be
called as a non-minimal $c<1$ non-critical string. Compared with
the standard minimal model case, we can allow irrational values of
the central charge. After the Lorentz transformation this
background is equivalent to the usual $c=1$ string with a
non-standard and time-dependent Liouville potential. We identified
the corresponding time-dependent fermi surface in the dual $c=1$
matrix model. We compared the tree level scattering S-matrices in
the matrix model with those computed in 2d string theory and found
a perfect agreement. It would be interesting to find a precise
matrix model dual description beyond the tree level. Notice that
in other words, we have discussed how to realize the Lorentz
transformation, which is only manifest in 2d closed string theory,
 from the viewpoint of its holographic dual
 open string theory defined in the lower dimension. We also
proposed an equivalent topological string description on a series
of specific non-compact Calabi-Yau manifolds given by \groundring.

Another interesting quantity in the matrix model which we may be
able to compare with string theory will be the macroscopic loop
operator $\log (\mu_B+\Phi)$ \Mac\ (see also the review \MaR). As
shown in \Ma, it is equivalent to the FZZT-brane with Dirichlet
boundary condition in the time-direction when we assume the usual
vacuum $b=1$. For generic $b$, however, this does not look
straightforward because there is a linear dilaton in the time
direction and its Dirichlet boundary condition is not
well-defined. Since the loop operator itself is well defined even
in time-dependent background, this will be an intriguing future
problem\foot{For example, we can write down the deformation of
fermi surface due to the loop operator as in \TaD. Though we can
read off from this the one-point function of the corresponding
D-brane boundary state, its explicit form does not look so simple
except the ordinary case $b=1$.}. A related question will be the
D-brane spectrum\foot{ In the minimal model case, we can associate
the moduli of FZZT-branes with a Riemann-surface in non-compact
Calabi-Yau spaces  \ADKMV \SH\ at tree level.} in the dual
non-compact Calabi-Yau \groundring\ and its relation to the
boundary states in our background \tldaction.

We also noticed that the matrix model description predicts a
series of new boundary states in our backgrounds of two
dimensional string theory. This is a generalization of the known
boundary states for the rolling tachyon
$T(t)\sim \cosh(t)$ \Sen\ in
our time-like linear dilaton case.

Furthermore, we discussed the 2d string theory whose matter part
(or time part) is given by a time-like Liouville theory. We
considered the dual matrix model configuration. Interestingly, we
noticed that the dual cosmological constant does not automatically
appear when the original cosmological constant is
non-zero\foot{This may also solve a similar puzzle in the
$SL(2,R)/U(1)$ WZW model at the level $0<k<2$ and its
sine-Liouville dual noticed in \HT.}. Also we find that the ground
ring structure looks very similar to that of the minimial string.
To understand better the duality between the time-like Liouville
theory and our matrix model background, we will need to compare
dynamical quantities like scattering amplitudes. Even though it is
not clear if we can define scattering processes in the string
theory side of the time-like Liouville theory, it seems that we
can consider an incoming wave in the matrix model background and
try to follow the time-evolution. This issue will also deserve a
future study.

\centerline{\bf Acknowledgments}

I am very grateful to M. Aganagic, D. Gaiotto, S. Gukov, T.
Harmark, Y. Hikida, N. Itzhaki, J. Karczmarek, D. Kutasov, J.
McGreevy, N. Saulina, D. Shih, A. Strominger, C. Vafa, E.
Verlinde, and S. Yamaguchi for useful comments and discussions.
This work was supported in part by DOE grant DE-FG02-91ER40654.

\appendix{A}{Comparison of S-matrices in 2D String Theory}

Here we summarize the results of S-matrix in 2d String Theory.
First we follow the notation of \DK\ i.e. $\ap'=2$ and the
Liouville potential is $\mu \int d\sigma^2 e^{-\s{2}b\phi}$. The
vertex operators are given\foot{Here we have shifted $k$ in \DK\
by $\ap=0$.} by $e^{ik X+(-Q/\s{2}+|k|)\phi}\ \ (Q=b+1/b)$ and $X$
is now Euclidean. We define the `leg factor' (see \emission)
\eqn\legdk{
\Delta(k)=\f{\Gamma(1-\f{\s{2}k}{b})}{\Gamma(\f{\s{2}k}{b})} \ \
(k>0),\ \ \ \ \Delta(k)=\f{\Gamma(1+\s{2}bk)}{\Gamma(-\s{2}bk)}\ \
(k<0).} The S-matrix of three particles are given by \eqn\threeps{
S^{(3)}(k_1,k_2,k_3)=\delta(k_1+k_2+k_3+q/\s{2})\cdot
(\mu\gamma(b^2))^{s}\cdot\prod_{i=1}^3 \left(-\pi
\cdot\Delta(k_i)\right), } where $s$ is the number of insertions
of the Liouville potential term so that it satisfies the momentum
conservation \eqn\momeone{
\sum_{i=1}^3|k_i|-\s{2}bs=\f{1}{\s{2}}Q.} The four point function
is \eqn\fourteps{\eqalign{
&S^{(4)}(k_1,k_2,k_3,k_3)=\delta(k_1+k_2+k_3+k_4+\s{2}q)\cdot
(\mu\gamma(b^2))^{s}\cdot\prod_{i=1}^4 \left(-\pi
\cdot\Delta(k_i)\right)\cr &\ \ \cdot \left[\f{1}{\s{2}b}\left(
|k_1+k_2+q/\s{2}|+
|k_1+k_3+q/\s{2}|+|k_1+k_4+q/\s{2}|\right)-\f{1+b^2}{2b^2}\right],}
} where $s$ is given by \eqn\mometow{
\sum_{i=1}^4|k_i|-\s{2}bs=\s{2}Q.} Let us compare these results of
$1\to 2$ and $1\to 3$ scattering with those in the matrix model
computed in section 3. To match the convention we have to return
to the Minkowski signature with $\al=1$ unit performing the
scaling \eqn\scaleone{\s{2}k\to -iE.} Then they are written as
follows\foot{Here we neglect the common factor $\s{2}$ which comes
from the delta function normalization. Also we put a factor
$\f{1}{b}$ due to the integration over the zero mode of $\phi$,
which is not explicitly written in \DK.} \eqn\smatst{\eqalign{
S^{(3)}_{1\to 2}&=\f{i}{b}\cdot\delta(E_1+E_2+E_3+iq)\cdot
(\mu\gamma(b^2))^{-1-iE_1 /b}\cdot (-\pi)^3 \cr &\ \ \ \times
\f{\Gamma(1+iE_1/b)}{\Gamma(-iE_1/b)}\f{\Gamma(1-ibE_2)}
{\Gamma(ibE_2)}
\f{\Gamma(1-ibE_3)}{\Gamma(ibE_3)}\cr
&=\delta(E_1+E_2+E_3+iq)\cdot (\mu\gamma(b^2))^{-1-iE_1 /b}\cdot
(-\pi)^3\cdot E_1E_2E_3 \cr &\ \ \
\times\f{\Gamma(iE_1/b)}{\Gamma(-iE_1/b)}
\f{\Gamma(-ibE_2)}{\Gamma(ibE_2)}
\f{\Gamma(-ibE_3)}{\Gamma(ibE_3)}, \cr  S^{(4)}_{1\to
3}&=\f{i}{b}\cdot\delta(E_1+E_2+E_3+E_4+2iq)\cdot
(\mu\gamma(b^2))^{-2-iE_1 /b}\cdot(-\pi)^4\cdot (-1-iE_1/b) \cr &\
\ \ \times
\f{\Gamma(1+iE_1/b)}{\Gamma(-iE_1/b)}\f{\Gamma(1-ibE_2)}
{\Gamma(ibE_2)}
\f{\Gamma(1-ibE_3)}{\Gamma(ibE_3)}\f{\Gamma(1-ibE_4)}
{\Gamma(ibE_4)}\cr
&=\delta(E_1+E_2+E_3+E_4+2iq)\cdot (\mu\gamma(b^2))^{-2-iE_1
/b}\cdot(-\pi)^4\cdot (ib-E_1)E_1E_2E_3E_4 \cr &\ \ \ \times
\f{\Gamma(iE_1/b)}{\Gamma(-iE_1/b)}
\f{\Gamma(-ibE_2)}{\Gamma(ibE_2)}
\f{\Gamma(-ibE_3)}{\Gamma(ibE_3)}\f{\Gamma(-ibE_4)}
{\Gamma(ibE_4)}.}}
In the end we can show that the string theory S-matrices exactly
agree with those of $c=1$ matrix model \smatrixtld\ taking into account the
scaling $\mu\to\mu\gamma(b^2)$ and the field normalization
\eqn\fieldre{S_{mat}(t,\phi) =\left(-\f{i}{2\pi}\right)\cdot
S_{string,\al=1}(t,\phi).}

\appendix{B}{Computation of Free Energy}

It will also be useful to find the tree level free energy in the
matrix model background. We can estimate the expectation values
$v_{n,m}$ of the (classical) $w_{\infty}$ algebra \AJ \PoR. Since
the fermi sea extends infinitely, we need a cut off $|x|<\Lambda$.
We can explicitly evaluate the classical contributions in the late
time $t>>1$ as follows
 \eqn\winft{\eqalign{
v_{n,m}&\equiv e^{(n-m)t}\int_{F-F_0}
\f{dxdp}{2\pi}(-p-x)^{m}(p-x)^n \cr &=-\f{e^{(n-m)t}}{m+1}\left[
\int^{\Lambda}_{a}\f{dx}{2\pi}\left(2\mu^{1/b^2}e^{(1-1/b^2)\ti{t}}
|x|^{-1/b^2}\right)^{m+1}\cdot (2|x|)^n\right]\cr
&-\f{e^{(n-m)t}}{n+1}\left[
\int^{\Lambda}_{a}\f{dx}{2\pi}\left(2\mu e^{(b^2-1)\ti{t}}
|x|^{-b^2}\right)^{n+1}\cdot (2|x|)^m\right]\cr &=C_{n,m}\cdot
\mu^{\f{n+m+2}{1+b^2}}\cdot
e^{\left[\f{2b^2}{1+b^2}(n+1)-\f{2(m+1)}{1+b^2}\right]\ti{t}}
+(\Lambda\  dependent\  term),}} where $F$ is our fermi surface
and $F_0$ is the one defined by $p^2-x^2=0$. $a$ is defined to be
$a=\f{1}{2}\cdot b^{\f{2}{1+b^2}}(1+1/b^2)$. The constant $C$ is
given by \eqn\const{C_{n,m}=\f{2^{n+m+1}}{\pi((n+1)b^2-m+1)}\cdot
\left( \f{b^2}{m+1} \cdot (2a)^{n+1-(m+1)/b^2} -\f{1}{n+1} \cdot
(2a)^{-(n+1)b^2+m+1}\right).} In particular, the energy of the
system is \eqn\energym{v_{1,1}=C_{1,1}\cdot
\mu^{\f{4}{1+b^2}}\cdot e^{4\f{b^2-1}{b^2+1}\ti{t}}.}
 On the other hand, in the string theory on the background
 \tldaction\ we can estimate the time-dependent energy
\eqn\freet{F(t)\sim (g_s)^{-2}=e^{-2qt-2Q\phi_0},} where $\phi_0$
is the characteristic value given by $\mu e^{2b\phi_0}=1$. Then we
can see that the both results \energym\ and \freet\ agree with
each other because \eqn\agree{\mu^{\f{4}{1+b^2}}\cdot
e^{4\f{b^2-1}{b^2+1}\ti{t}}=\mu^{\f{4}{1+b^2}}\cdot
e^{-2qt-2\f{q^2}{Q}\phi_0}=e^{-2qt-2Q\phi_0}.}

\listrefs

\end